\documentclass[]{emulateapj}

\def\Mpc{\ {\rm Mpc}}

\def\kms{{\ }{\rm km}\,{\rm s}^{-1}}
\def\yr{{\rm yr}}
\def\Msun{M_{\odot}}
\def\Mstel{M_\ast}
\def\Mdisk{M_{\rm \ast, disk}}

\def\resp{respectively}
\def\sersic{S\'ersic }
\def\ns{n_{\rm S\acute{e}rsic}}
\def\to{$\to$\ }
\def\bfr{\bf\color{red}}

\def\ssfr{{\rm sSFR}}%_{tot}}}
\def\ssfrd{{\rm sSFR_{disk}}}

\def\sfr{{\rm SFR}}
\def\bof{B04$_4$}
\def\bos{B04$_7$}
\def\MS{MS}
\def\sigms{\sigma_{\rm MS}}
\def\betare{\beta_{\rm disk}}
\def\betacon{\beta_{\rm control}}

\def\betaconval{-0.05\pm0.04}

\mathchardef\mhyphen="2D

\def\cite{{\bfr CITE}}
\usepackage{MnSymbol}
\usepackage{wasysym}
\usepackage{latexsym}
\usepackage{amssymb}
\usepackage{graphicx}
\usepackage{mathrsfs}
\usepackage{amsmath}
\usepackage[bottom]{footmisc}
\usepackage{color}
\usepackage{hyperref}
\shortauthors{ABRAMSON ET AL.}
\shorttitle{SPECIFIC STAR FORMATION RATES OF DISKS}

\begin{document}

\title{The Mass-Independence of Specific Star Formation Rates in Galactic Disks}

\slugcomment{Accepted to ApJ Letters}

\author{
Louis E. Abramson\altaffilmark{1,2,$\ast$}, 
Daniel D. Kelson\altaffilmark{2}, 
Alan Dressler\altaffilmark{2}, 
Bianca Poggianti\altaffilmark{3},
Michael D. Gladders\altaffilmark{1},
Augustus Oemler, Jr\altaffilmark{2},
and Benedetta Vulcani\altaffilmark{4}
}

%------------------------------------------------------------------------------------------------------------------------------------------
%------------------------------------------------------------------------------------------------------------------------------------------

\begin{abstract}
The slope of the star formation rate/stellar mass relation (the SFR ``Main Sequence"; $\sfr$--$\Mstel$) is not quite unity: specific star formation rates ($\sfr/\Mstel$) are weakly-but-significantly anti-correlated with $\Mstel$. Here we demonstrate that this trend may simply reflect the well-known increase in bulge mass-fractions -- portions of a galaxy {\it not} forming stars -- with $\Mstel$.  Using a large set of bulge/disk decompositions and SFR estimates derived from the Sloan Digital Sky Survey, we show that re-normalizing SFR by {\it disk} stellar mass ($\ssfrd\equiv\sfr/\Mdisk$) reduces the $\Mstel$-dependence of SF efficiency by $\sim0.25$ dex per dex, erasing it entirely in some subsamples. Quantitatively, we find $\log\ssfrd$--$\log\Mstel$ to have a slope $\betare\in[-0.20,0.00]\pm0.02$ (depending on $\sfr$ estimator and Main Sequence definition) for star-forming galaxies with $\Mstel\geq10^{10}\Msun$ and bulge mass-fractions $B/T\lesssim0.6$, generally consistent with a pure-disk control sample ($\betacon=\betaconval$). That $\langle\sfr/\Mdisk\rangle$ is (largely) independent of host mass for star-forming disks has strong implications for aspects of galaxy evolution inferred from any $\sfr$--$\Mstel$ relation, including: manifestations of ``mass quenching" (bulge growth), factors shaping the star-forming stellar mass function (uniform $d\log\Mstel/dt$ for low-mass, disk-dominated galaxies), and diversity in star formation histories (dispersion in $\sfr(\Mstel,t)$).  Our results emphasize the need to treat galaxies as composite systems -- not integrated masses -- in observational and theoretical work.
\end{abstract}

%\keywords{
%galaxies: evolution
%}

\altaffiltext{1}{
Department of Astronomy \& Astrophysics and Kavli Institute for Cosmological Physics, The University of Chicago, 5640 S Ellis Ave, Chicago, IL 60637, USA
}
\altaffiltext{2}{
The Observatories of the Carnegie Institution for Science, 813 Santa Barbara St, Pasadena, CA 91101, USA
}
\altaffiltext{3}{
INAF-Osservatorio Astronomico di Padova, Vicolo Osservatorio 5, 35122 Padova, Italy
}
\altaffiltext{4}{
Kavli Institute for the Physics and Mathematics of the Universe (WPI), Todai Institutes for Advanced Study, University of Tokyo, Kashiwa 277-8582, Japan
}
\altaffiltext{$\ast$}{
\href{mailto:labramson@uchicago.edu}{\tt labramson@uchicago.edu}
}

%------------------------------------------------------------------------------------------------------------------------------------------
%------------------------------------------------------------------------------------------------------------------------------------------

\section{Introduction}
\label{sec:intro}

The observation of a correlation between galaxy star formation rates (SFRs) and stellar masses ($\Mstel$) has generated considerable interest. Seen from $z = 0$ to $z > 2$ \citep[e.g.,][and references therein]{Brinchmann04, Daddi07, Wuyts11, Guo13}, this SFR ``Main Sequence" (\MS) may encode fundamental information about galaxy evolution.

Uncontroversial is the fact that the \MS\ has fallen monotonically since at least $z\sim2$ \citep[e.g.,][]{Noeske07, Rodighiero10, Whitaker12}.  Observed at all $\Mstel\gtrsim10^{10}\Msun$, this phenomenon must contribute significantly to the precipitous decline in cosmic star formation seen over the same epoch \citep[e.g.,][]{Lilly96, Madau96, Cucciati12}.

However, while its gross evolution is increasingly well-understood, the slope and dispersion of the \MS\ remain uncertain. Such uncertainty arises (at least) from dependencies on SFR indicators \citep[e.g.,][Figure 4]{Pannella09}, the definition of ``star forming" \citep[e.g.,][\S7.5]{Salim07}, and a lack of high-redshift data at moderate-to-low $\Mstel$ \citep[e.g.,][Figure 1]{Whitaker12}.

Despite these issues, if its evolution reflects that of individual systems, the slope and dispersion of the \MS, their time-dependence, and their interpretation have deep implications for pictures of galaxy growth. With the \MS\ broadly reproducible in cosmological simulations \citep[][]{Keres05, Neistein08, Lagos11, Hopkins13} and actively employed as a basis/constraint for evolutionary models \citep[][]{PengLilly10, Leitner12, Behroozi13}, understanding such details is increasingly important.

Here we reinterpret the slope of the \MS.

% --- TABLE 1 ---
\begin{deluxetable*}{lccl}
\tablecolumns{4}
\tablecaption{Quantities}
\tablehead{
\colhead{Quantity} &
\colhead{Unit} &
\colhead{Source\tablenotemark{a}} &
\colhead{Definition}
}
\startdata
$r_{\rm disk}$		& SDSS mag\tablenotemark{b}	& 1 & Disk absolute $r$ magnitude $k$-corrected to $z=0.1$\\
$(g-r)_{(\rm disk)}$	& SDSS mag\tablenotemark{b}	& 1 & (Disk) rest-frame color $k$-corrected to $z=0.1$\\
$z$				& $\cdots$					& 1 & Galaxy redshift\\
$\ns$			& $\cdots$					& 1 & Global $r$-band \sersic index\\
$b/a$			& $\cdots$					& 1 & Global $r$-band axis ratio (1 - ellipicity)\\
$M_{\ast(,\,\rm disk)}$& $\Msun$				& 2(3) & (Disk) stellar mass\\
SFR 				& $\Msun\, {\rm yr^{-1}}$ 		& 2 & Aperture-corrected star formation rate (median of PDF)\\
sSFR 				& ${\rm yr^{-1}}$ 			& 2 & Galaxy specific star formation rate ($\sfr/\Mstel$) \\
$\ssfrd$			& ${\rm yr^{-1}}$ 			& 3 & Disk-mass-normalized star formation rate ($\sfr/\Mdisk$)
\enddata
\tablenotetext{a}{1--S11; 2--\bof/\bos; 3--Derived.}
\tablenotetext{b}{AB system offsets are $<0.01$ mag.}
\label{tbl:params}
\end{deluxetable*}

The \MS\ is conveniently recast in terms of galaxies' {\it specific} star formation rates -- $\ssfr\equiv\sfr/\Mstel$ -- or fractional mass-growth per unit time. If constant, $\ssfr$ is the (inverse) $\Mstel$ $e$-folding timescale.

The $\Mstel$-dependence of sSFR -- the departure of the \MS\ slope from unity -- contains information about the ``efficiency" of SF across the galaxy mass spectrum.\footnote[5]{More direct definitions of ``SF efficiency" relate $\sfr$ to a gas mass, but $\ssfr$ is {\it an} efficiency metric.}  Typically, it is parametrized by the power-law index: 
\begin{equation}
	\beta\equiv\frac{d\log\ssfr}{d\log\Mstel}.
\end{equation}

If all galaxies formed stars with equal efficiency, $\beta$ would be identically zero.  Observationally, $\beta$ appears {\it close} to zero, permitting convenient approximations in evolutionary models \citep[e.g.,][]{PengLilly10}; $\ssfr(t)$ is {\it nearly} independent of mass, so the entire star-forming population is {\it nearly} describable by a single number (absent significant dispersion at fixed $\Mstel$; see Section \ref{sec:scatter}). 

Yet, $\beta$ is {\it not} zero. Many studies using SFR indicators from the UV through the radio have concluded that, above $10^{10}\Msun$, $-0.6\lesssim\beta\lesssim-0.1$ for $z\lesssim2$ \citep[][but cf.\ \citealt{Pannella09}]{Brinchmann04, Salim07, Karim11, Whitaker12}. \citet[][]{PengLilly10} and \citet[][]{Whitaker12} find $\beta\simeq0$ for blue galaxies (see Section \ref{sec:implications} below), but that $\beta$ is significantly negative for the global star-forming population seems secure.

The implication of $\beta<0$ is that low-mass galaxies grow (logarithmically) faster than higher-mass contemporaries.  Interesting on its own, this fact is important also because $\beta$ informs two other key questions: Why has the shape of the star-forming stellar mass function remained unchanged since $z\sim2$ \citep[e.g.,][]{Ilbert10, Tomczak14}? What {\it stops} star formation? 

Setting aside the mass function for now (see Section \ref{sec:implications} and the extensive treatment of \citealt{PengLilly10}) the question of what stops SF in galaxies nicely illustrates $\beta$'s influence on \MS-based evolutionary models.

In the $\beta\rightarrow0$ limit, galaxy evolution is binary: systems are either star-forming -- growing in lock-step with all other such objects -- or not.  An implication is that mechanisms taking galaxies from the first population into the second act quickly and operate across all $\Mstel$.

Conversely, if $\beta$ is substantially negative (as is likely), galaxy evolution is more nuanced.  A galaxy's global SF efficiency changes as it grows, gradually falling to negligible levels with time. ``Quenching" is thus a mix of processes pulling systems vertically off the \MS\ and lowering $\ssfr$s as they move along it ($\beta$ reflects the latter).

Many mechanisms have been proposed that implicitly or explicitly account for mass-dependent $\ssfr$s, including virial-heating of the circumgalactic medium by dark matter halos (inducing ``hot-mode" accretion) and AGN activity \citep[e.g.,][]{Dekel04, Keres05, Croton08, vandeVoort11b}.  Such processes may be at work, but they are not directly coupled to the observables in $\ssfr$--$\Mstel$, so hypotheses are complicated by uncertainties in linking these phenomena.

Indeed, observationally, there is a deeper concern. $\ssfr=\sfr/\Mstel$ (hence $\beta$) is biased, {\it prima facie}, as a description of SF as the numerator has essentially nothing to do with a significant part of the denominator -- the bulge. Given the well-known correlation of bulge mass-fractions, $B/T$, with $\Mstel$, $\beta<0$ is expected simply because ever smaller portions of a galaxy participate in SF, independent of the nature of the SF itself.

If $\ssfr$--$\Mstel$ is to add meaningfully to our knowledge of galaxy evolution, at a minimum, the extent to which $\beta$ reflects changes in the quality of SF ({\it how}) versus the proportion of a galaxy contributing to it ({\it where}) must be understood.  Large spectrophotometric surveys -- such as the Sloan Digital Sky Survey \citep[SDSS --][]{York00} -- enable this.  

Below, we demonstrate the importance of recognizing {\it where} SF occurs, showing that most-to-all of $\beta$ can be erased simply by redefining ``$\ssfr$" using the mass in galactic disks.

%------------------------------------------------------------------------------------------------------------------------------------------
%------------------------------------------------------------------------------------------------------------------------------------------

\section{Data}
\label{sec:data}

We use data from the Seventh SDSS Data Release \citep[DR7 --][]{SDSS_DR7}, drawing $\sfr$s and $\Mstel$ from \citet[][hereafter B04]{Brinchmann04},\footnote{\url{www.mpa-garching.mpg.de/SDSS/DR7/sfrs.html}} and 2D bulge/disk decompositions from \citet[][hereafter S11]{Simard11}. Given their extensive past use, however, we analyze DR4-based B04 data -- based on a different $\sfr$ calculation\footnote{Known to overestimate $\sfr$ in quiescent galaxies; \url{www.mpa-garching.mpg.de/SDSS/DR4/Data/sfr_catalogue.html}.} -- in paralel. Below, \bof\ and \bos\ refer \resp\ to DR4-/DR7-based measurements while ``B04" refers to the original paper (\bos\ lacks a stand-alone reference at present).

Both SFR and $\Mstel$ assume a \citet{Kroupa01} initial mass function. B04 give these quantities as probability distributions. We adopt the median total values, but results are unchanged if the mean or mode is used instead. 

Below, we denote quantities describing disks by the (additional) subscript ``$_{\rm disk}$". Quantities lacking this tag describe global galaxy properties.  Table \ref{tbl:params} lists all parameters and their sources; $(H_0,\Omega_m,\Omega_\Lambda)=(70\kms\Mpc^{-1}, 0.3, 0.7)$ is assumed everywhere.

\begin{figure*}[t!]
\centering
\includegraphics[width = 1.7\columnwidth, trim = 0cm 0.5cm 0cm 1cm]{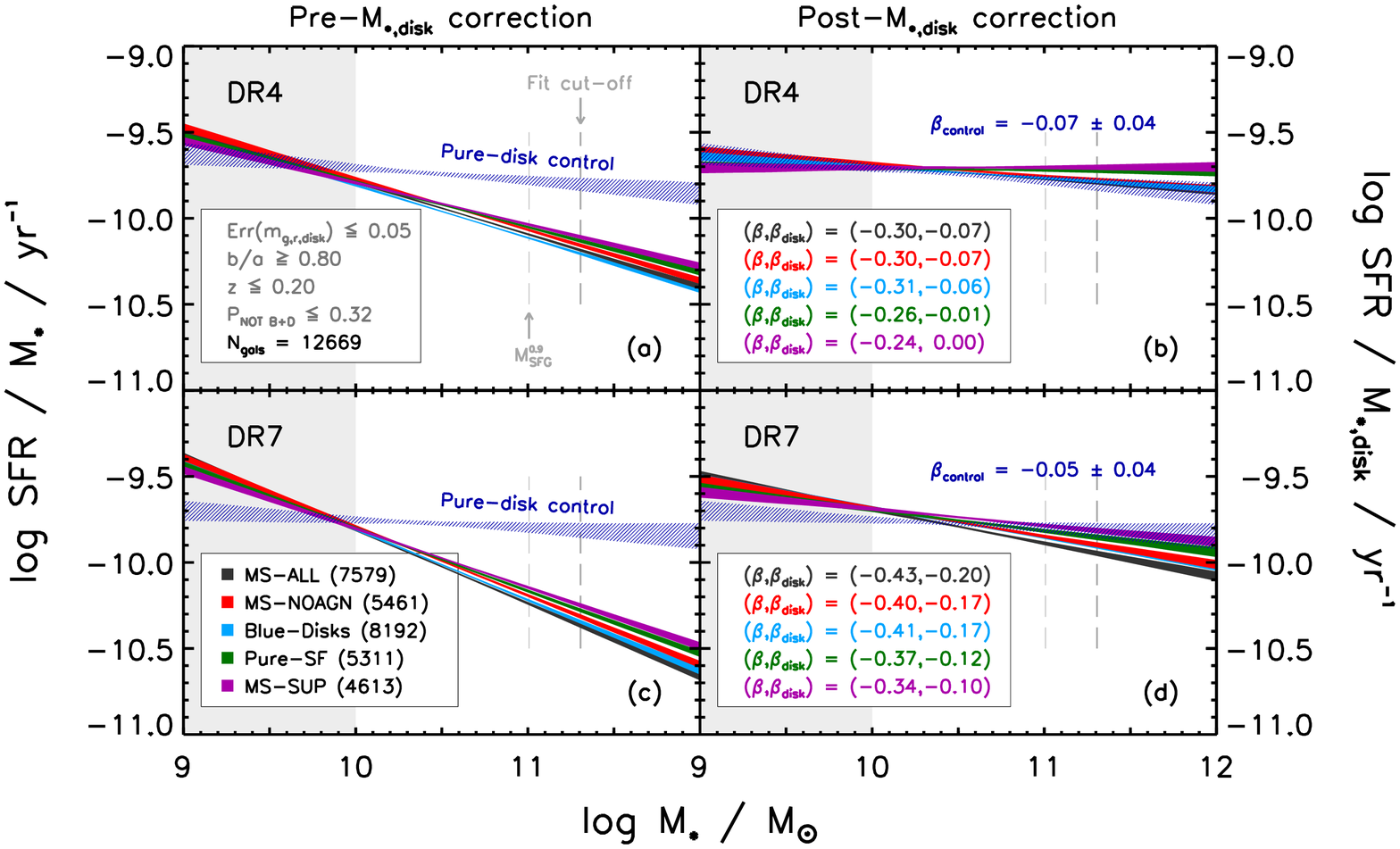}
\caption{\MS\ fits before/after $\Mdisk$ re-normalization ({\it left}/{\it right}) using \bof/\bos\ data ({\it top}/{\it bottom}). $\sfr/\Mstel$-- and $\sfr/\Mdisk$--$\Mstel$ have slopes $\beta,\, \betare$, \resp. Pure-disk control $\sfr/\Mstel$--$\Mstel$ is also plotted (blue hatching). Grey denotes regions of possible $\ssfr$ bias; only data at $10^{10}\Msun\leq\Mstel\leq 2\times M_{\rm SFG}^{0.9}$ (the 90$^{\rm th}\, \Mstel$-percentile for pure-SF galaxies) were fit. Band widths denote 1-$\sigma$ uncertainties.}
\label{fig:fits}
\end{figure*}

\subsection{Bulge/Disk Decompositions}

We use S11 ``fixed $n_b$" fits, where $n_b\equiv 4$ is the \sersic index of the bulge component.  These are appropriate for almost all sources (S11 \S4.2), but results are qualitatively unaffected if the ``free $n_b$" models are used instead. We take disk and total $g,\,r$ absolute magnitudes from the fits. Employing model-independent Petrosian magnitudes from the NYU Value Added Galaxy Catalog \citep[][]{Blanton05VAGC} does little but reduce $\Mdisk$ for blue disks (Section \ref{sec:systematics}).

To avoid dust and $S/N$ effects, we limit our analysis to face-on galaxies ($b/a\geq0.8$) with well-measured disk fluxes (Err$(g,r)_{\rm disk}\leq0.05$) and total masses ($\Mstel\geq10^9\Msun$). We further restrict the \MS\ samples (see Section \ref{sec:results}) to galaxies requiring a two-component bulge$+$disk model.\footnote{$P$(NOT 2-component) $<0.32$; S11 \S4.2.1} Relaxing these cuts affects $\beta_{\rm (disk)}$ less than other systematics, but employing them ensures maximally accurate $\Mdisk$, $\sfr$, and meaningful $\Mdisk$ corrections.

\subsection{The Sample}
In total, 12669 systems common to DR4/DR7 meet these criteria, with median $(z, \Mstel) = (0.08,5.5\times10^{10}\Msun)$. These include: ``Pure-SF" (42\%); ``SF/AGN composite" (9\%); ``AGN" (6\%); ``LINER" (15\%); and ``Unclassifiable" galaxies (no detected emission; 28\%). 

Given the SDSS spectroscopic limit, this sample is roughly complete to $4\times10^{10}\Msun$ for star-forming systems (assuming $90^{\rm th}$-percentile color and redshift). However, $\sfr$ completeness -- set by line-flux, spectral $S/N$, and broadband colors -- is of greater concern since it can distort fits in the $\ssfr$--$\Mstel$ plane. Since photometric completeness is not an issue and $\sfr\approx 1\,\Msun\,\yr^{-1}$ is well-measured by B04, the data should be relatively unbiased above the corresponding \MS\ mass, $\Mstel\approx10^{10}\Msun$. We perform all fits above this limit and derive statistics using $1/V_{\rm max}$ weighting.

Typical half-light radii are $\approx3\farcs8$. As $\langle{\rm FWHM_{SDSS}}\rangle\approx1\farcs4$, disks should be well-resolved.

\subsection{Calculation of Disk Masses}

We estimate $\Mdisk$ empirically.  First, we select a sample of disk-dominated systems -- bulge-to-total flux ratio $(B/T)_r\leq0.2$ -- whose color and mass should largely reflect those of pure disks.  We then calculate $r$-band mass-to-light ratios, $\Upsilon_r\equiv\Mstel/L_r$, and derive $\langle\log\Upsilon_r(g-r)\rangle$ by fitting a second-order polynomial. We do this independently for \bof\ ($\Mstel$ from spectral fitting by \citealt{Kauffmann03}) and \bos\ ($\Mstel$ from SED fitting). \bos\ yields $\langle\log\Upsilon_r(g-r)\rangle=-1.00+2.47(g-r)-0.85(g-r)^2$, with \bof\ offsets $\lesssim0.06$ dex for $g-r<0.71$ ($90^{\rm th}$-percentile disk color). Using $g,\,r$ absolute disk magnitudes:
\begin{multline}
	\label{eqn:mdisk}
	\log\Mdisk/\Msun= \\ -0.4(r_{\rm disk}-r_\odot)+\langle\log\Upsilon_r(g-r)_{\rm disk}\rangle,
\end{multline}
where $r_\odot=4.64$ \citep{BlantonRoweis07}.

We then define:
\begin{equation}
	\ssfrd\equiv\sfr/\Mdisk.
\end{equation}
This may not {\it formally} correspond to ``the sSFR of the disk" as bulge/nuclear regions may contribute some SF, but to ease discussion and because such contributions should be small, we use ``$\ssfrd$" instead of ``$\Mdisk$-normalized SFR" below. 

\begin{figure*}[t!]
\centering
\includegraphics[width = 1.4\columnwidth, trim = 1cm 1.5cm 1cm 2cm]{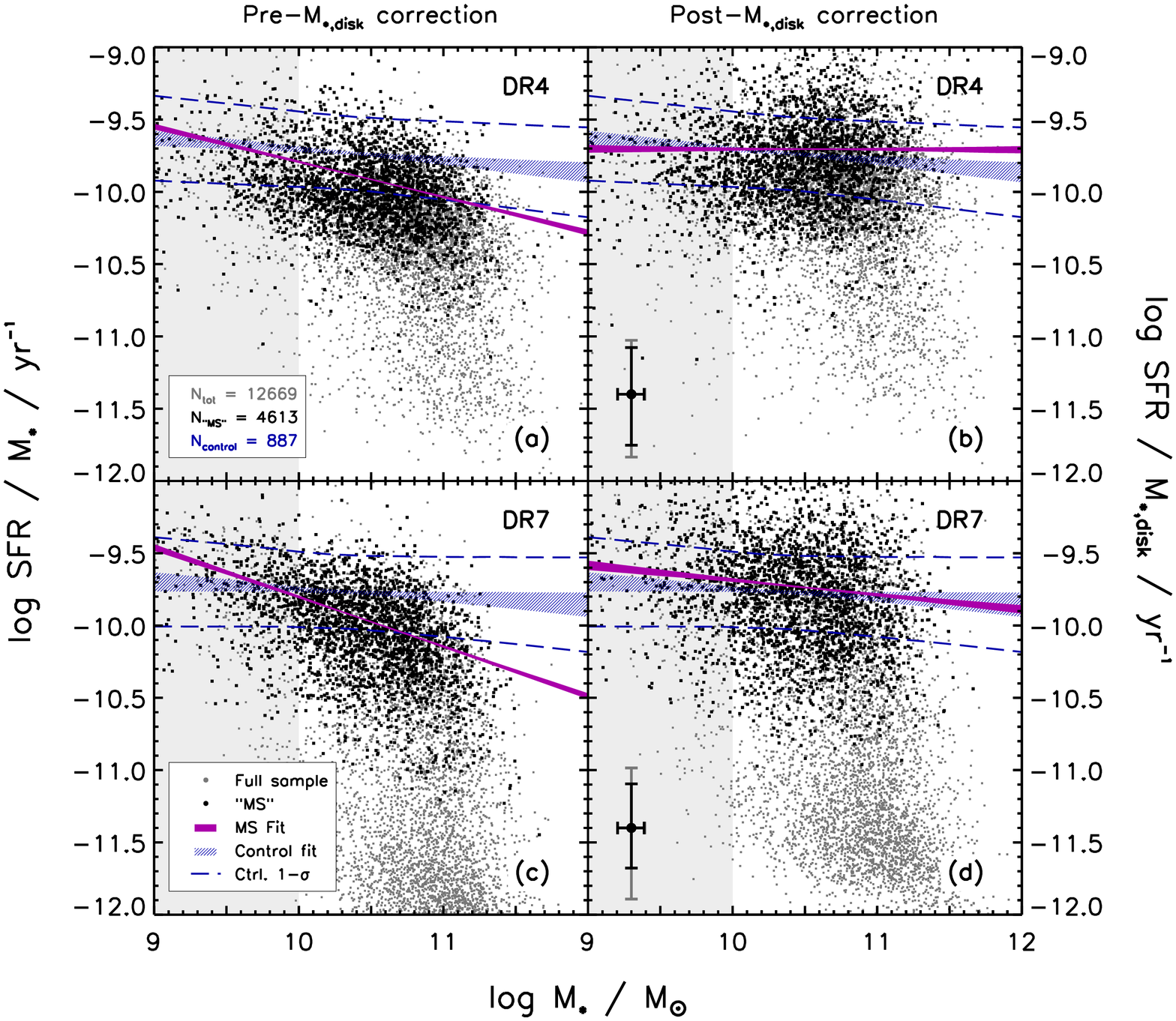}
\caption{As Figure \ref{fig:fits}, but showing data. Grey points are all galaxies, black the MS-SUPER sample. Fits are replotted from Figure \ref{fig:fits}. Dashes show 1-$\sigma$ control data spread (points not plotted). Control and \MS\ distributions agree well after $\Mdisk$ re-normalization.}
\label{fig:simple}
\end{figure*}

Median $1$-$\sigma$ uncertainty in $\Mstel$ using \bof\ or \bos\ is 0.09 dex. Scatter in $\log\Upsilon_r(g-r)$ is 0.12/0.08 dex, \resp.  Quality cuts ensure Err$(g-r)_{\rm disk}\leq0.07$ mag (the median is 0.03), so random errors in $\Mstel$ and $\Mdisk$ should be comparable. Formal $1$-$\sigma$ uncertainties in $\sfr$ are $\sim 0.3$ dex (either estimate) and therefore dominate.

%------------------------------------------------------------------------------------------------------------------------------------------
%------------------------------------------------------------------------------------------------------------------------------------------

\section{Results}
\label{sec:results}

Figure \ref{fig:fits} summarizes our analysis. Here we plot fits to the \MS\ in both $\log\ssfr$--$\log\Mstel$ (left) and $\log\ssfrd$--$\log\Mstel$ space (right). Because the locus has no formal definition, we approximate the \MS\ in 5 (non-independent) ways:

\begin{itemize}
	\item {\bf MS-ALL:} All galaxies with $\ssfr$ above \MS $-3\sigma$ (defined using \bos).
	\item {\bf MS-NOAGN:} The same, excluding AGN, Composite, and LINER galaxies.
	\item {\bf PURE-SF:} All pure-SF systems regardless of $\ssfr$; excludes AGN-contaminated and Unclassified galaxies.
	\item {\bf BLUE DISK:} All galaxies with $(g-r)_{\rm disk}\leq0.6$ regardless of spectral type or $\ssfr$.
	\item {\bf MS-SUPER:} Intersection of all of the above; the purest, but smallest, sample.
\end{itemize}

Also overplotted are results for a ``pure-disk control" sample (where $\ssfr=\ssfrd$) composed of pure-SF systems well-fit by a single-disk profile.\footnote{$P$(NOT 2-component)$\ \geq0.5$, $\ns\leq2$.}

Three points are clear: 
\begin{enumerate}
	\item The slope, $\beta$, of $\ssfr$--$\Mstel$ is substantially steeper for the \MS\ samples than for the pure-disk control (Figure \ref{fig:fits}{\it a,c}); 
	\item The slope, $\betacon$, of the pure-disk control is consistent with zero at the 1- to 2-$\sigma$ level (as seen at $z\sim1$ by \citealt{Salmi12});
	\item After $\Mdisk$ re-normalization, \MS\ slopes, $\betare$, and intercepts are similar to -- even consistent with -- those of the pure-disk controls (Figure \ref{fig:fits}{\it b,d}).
\end{enumerate}

Quantitatively, we find $-0.43\leq\beta\leq-0.24$ \citep[consistent with results from][]{Salim07, Karim11, Whitaker12}, but $-0.20 \leq\betare\leq0.00$. (Spreads reflect data set and inter-sample variations.) This $\sim0.25$ dex/dex enhancement is interesting in an absolute sense: it substantially (perhaps entirely) homogenizes mean SF efficiencies over more than a factor of 10 in $\Mstel$. But, it is the homogenization of galaxies spanning $0.1\lesssim B/T\lesssim0.6$ with pure disks ($\betacon=-0.05\pm0.04$) that suggests $\Mdisk$ re-normalization is physically meaningful.

Statistical uncertainties in $\beta$ and $\betare$ are $\sim0.02$, derived from fits to 100 bootstrap resamplings of the data at $10^{10}\Msun\leq\Mstel\leq2\times M^{0.9}_{\rm SFG}$ (90$^{\rm th}$-percentile $\Mstel$ for pure-SF galaxies). Systematics are clearly dominant, with \MS\ definition and $\sfr$ estimate both contributing at the $\Delta\betare\approx0.06$--0.10 level (Section \ref{sec:systematics}).

Figure \ref{fig:simple} shows the data. Grey points represent all galaxies, black the MS-SUPER sample, constituting $\sim 60\%$ of the SFR density in the local universe (MS-ALL comprises $\sim90\%$). Two additional points are illustrated here: 1) Dispersion in the \MS, $\sigms$, is substantial; 2) Pure disks move from the top of the $\ssfr$--$\Mstel$ relation to the middle of $\ssfrd$--$\Mstel$. We discuss $\sigms$ in Section \ref{sec:scatter}, but (2) is further evidence that the $\Mdisk$ correction is physically meaningful: not only is $\beta$ pushed close to $\betacon$, but the original \MS\ {\it distribution} is made to coincide with that of pure disks. Visually comparing the 1-$\sigma$ control spread (dashed blue lines) to that of $\ssfrd(\Mstel)$ emphasizes this point. 

In sum, re-normalizing $\sfr$ by $\Mdisk$ substantially (perhaps entirely) homogenizes SF efficiency in giant galaxies, placing bulge-dominated, $10^{11} \Msun$ systems near the level of pure disks one-tenth as massive.

\section{Systematics}
\label{sec:systematics}
%Though it has a similarly small effect on $\betare$, using $\Upsilon_r(g-r)$ from \citet{Bell03} boosts \bos-derived $\beta$ and $\betacon$ by $\sim 0.12$ dex
Once the \MS\ is defined -- itself a $\Delta\beta_{\rm(disk)}\sim0.1$ effect (Figure \ref{fig:fits}) -- two systematics affect $\betare$: $\Mdisk$ calculation and $\sfr$ estimation. 

%We caution that this relation over-predicts $\Upsilon_r(g-r\leq0.4)$ by $\gtrsim0.1$ dex compared to more modern SED fitting, however, and therefore disfavor its use.
$\Mdisk$ is affected by bulge/disk decomposition and $\Upsilon_r$ calibration. Using \bof\ or \bos\ masses to calibrate $\Upsilon_r$ has an effect comparable to statistical uncertainties. Adopting $\Upsilon_r(g-r)$ from \citet{Bell03} changes $\betare$ similarly, but can boost $\beta,\, \betacon$ by $\sim0.1$ (\bos\ $\sfr$s). Using Petrosian magnitudes can induce $\Delta\betare=0.08$ (both data sets), but only for the BLUE DISK (and thus MS-SUPER) samples. Comparing S11-based $\Mdisk$ to estimates derived from decompositions by \citet[][SDSS-based, but more complex than S11; $N_{\rm gals}=529$]{Gadotti09} or \citet[][fit to independent Millennium Galaxy Catalogue imaging  (\citealt{Liske03}); $N_{\rm gals}=770$]{Allen06}, we find no trends larger than the scatter ($\sim0.25$ dex) at $\Mstel\geq10^{10}\Msun$. Hence, $\sfr$ systematics likely drive uncertainty in $\betare$.

Figure \ref{fig:simple}{\it a,c} illustrates this. The (substantial) changes between \bof\ and \bos\ -- bi-modality at high mass, increased dispersion -- mainly reflect revised aperture corrections introduced after \citet{Salim07} found \bof\ to overestimate $\ssfr$ in quiescent galaxies. Using a common $\Mdisk$, we find $\Delta\beta({\rm B04_4}-{\rm B04_7})\simeq0.10$ for all \MS\ samples. Swapping B04 $\sfr$s for optical emission line estimates from the Padova-Millennium Galaxy and Group Catalogue \citep[PM2GC --][requiring no color-based corrections]{Calvi11}, we find $\betare^{\rm PM2GC}=-0.18\pm0.08$ for galaxies with $(g-r)_{\rm disk}\leq0.6$, consistent with the analogous $\betare$ obtained from \bos. Hence, given the \bof/\bos\ offsets, systematics in $\betare$ are likely $\sim0.1$ once the \MS\ is defined.

%------------------------------------------------------------------------------------------------------------------------------------------
%------------------------------------------------------------------------------------------------------------------------------------------
\section{Implications}
\label{sec:implications}

We have identified a quantity that is roughly constant for star-forming galaxies at $\Mstel\geq10^{10}\Msun$: $\sfr/\Mdisk$. This implies that SF efficiency {\it in the disks of star-forming galaxies} (even bulge-dominated ones) is largely independent of global galaxy properties (e.g., halo mass). This is qualitatively different from (if anticipated by) findings regarding uniform $\sfr/\Mstel$ in disk-dominated galaxies \citep{Salmi12}, blue galaxies (likely {\it because} they are disk-dominated; see Section \ref{sec:intro}), and the correlation of $B/T$ with position on the \MS\ \citep[][]{Martig09,Williams10,Lang14,Omand14}, which our measurement of $\betacon\cong0$ supports.

Indeed, our results suggest that the suppression of SF efficiency with $\Mstel$ due to bulge-growth is mostly superficial, caused by the association of ``SF efficiency" with $\ssfr$ and the conflation of {\it where} and {\it how} SF occurs. That is, a key aspect of ``mass-quenching" is ``bulge-building", distinct from processes affecting SF {\it where it occurs}. In this we echo \citet{Kennicutt94}. 

Whether bulge-growth is predominantly secular \citep[converting dynamically ``cold" disk material through, e.g., bar-instabilities;][]{KK04} or merger-driven \citep[adding ``hot" bulge material through interactions;][]{ToomreToomre72} is beyond the scope of this paper, but future measurements of {\it disk} stellar mass functions -- or indeed $\beta(\Mstel,z)$ -- may shed light on this question.  Regardless, investigations of halo heating and/or AGN-powered quenching might focus on the narrower question of how these mechanisms build/maintain bulges in healthy disks.  (This and the previous point is refined in the next section.)
% for galaxies of different $B/T$

A third implication is worth noting. Since $z\sim2$, the low-$\Mstel$ slope of the star-forming stellar mass function has remained constant at $\alpha\approx-1.4$, yet $\beta<0$ is reported over the same interval almost universally (see Section \ref{sec:intro} for references).  These are inconsistent observations: $\beta<0$ implies $\alpha$ should steepen (dramatically) with time. Our results suggest that, at $\Mstel\geq10^{10}\Msun$ (where it is measured at $z>0$), $\beta$ largely reflects $B/T$. Extrapolations from this regime to lower-$\Mstel$ -- where star-forming galaxies are bulgeless -- may thus be inappropriate. If in fact $\beta\rightarrow0$ at lower mass  -- as results from \citet{Salim07}, \citet{Karim11}, and \citet{Whitaker12} also hint -- the \MS\ and $\alpha$ would be reconciled.

%Results from \citep{Salim07}, \citet{Karim11}, \citet{Salmi12}, and \citet{Whitaker12} hint that such a ``break-over" in $\beta$ indeed occurs.    but more definitive measurements are required.}
% $\Mstel$ is simply the wrong mass to use when defining $\beta$.  The uniform $\sfr/\Mdisk$ displayed by galaxies of any $\Mstel$ preserves $\alpha$ under the reasonable assumption that mass-growth is dominated by {\it in situ} SF.-- if shown to hold at $z\lesssim2$ -- 

\section{The Width of the SFR Main Sequence}
\label{sec:scatter}

So far, we have neglected dispersion in the \MS, $\sigms$. Given \bof\ data, this appears reasonable: $\sigms\lesssim0.3$ dex, consistent with formal errors (Figure \ref{fig:simple}).  However, \bos\ and numerous other data sets \citep[e.g.,][PM2GC]{Salim07,Oemler13a} suggest $\sigms\sim0.4$--0.6 dex (peak-to-peak $\Delta\ssfr(\Mstel)\gtrsim1$ order of magnitude), implying that the width of the \MS\ is qualitatively and quantitatively important.

Qualitatively, since $\beta_{\rm(disk)}\approx0$, $\sigms>0$ is necessary to preserve diversity in star formation histories (SFHs) as independently suggested by, e.g., stellar population synthesis (\citealt{Poggianti13a}; at least when $\Mstel(t)\approx\Mdisk(t)$). Comparing \citet[][Figure 19]{PengLilly10} with \citet[][Figure 2]{Gladders13b} reveals the contrast between $(\beta,\sigms)=(0,0)$ and $\neq(0,0)$, \resp, in terms of SFH diversity.%{\bfb This interpretation of $\sigms$ is also put forward by \citet{Guo13}.}

Inversely, real dispersion quantitatively complicates the determination of SFHs based on \MS\ evolution (Section \ref{sec:intro}): one must model $\sigms(\Mstel,t)$. How this could be done is unclear; data are scanty at $\Mstel\ll10^{10}\Msun$ and $z\gg1$ -- key parameter space when modeling Milky Way analogs -- and local measurements suggest $\sigms$ (and therefore its navigation) only becomes more important in this mass regime \citep[][\S7.5]{Salim07}.

Regardless, assuming it can be precisely measured, interpreting $\sigms$ will remain a challenge. Different $\sfr$ indicators probe different timescales ($\sim10^7$ vs.\ $10^8$--$10^9$ yr for optical and UV/IR metrics, \resp), so ambiguity in the causes of $\sigms(t)$ -- e.g., minor-mergers/starbursts \citep{Dressler13,Abramson13}, extended periods of enhanced gas accretion, stochasticity -- and thus its relevance to the ``fundamental" $\dot\Mstel$ history of galaxies may persist. If so, the utility of the \MS\ as a model for individual systems will remain questionable.

One can always imagine the opposite, however. If $\sigms$ is ``truly" small \citep[e.g.,][]{Salmi12}, our results suggest a quasi-identical SFH for all galactic disks (up to a scaling), with global galaxy-to-galaxy variations coming from bulge-building or environmental developments. Future IFU/resolved spectroscopic studies of galaxies at all redshifts could shed substantial light on this issue.

In sum, the ``$\Mdisk$ correction" is surely not the end of the story.  Though it homogenizes star-forming disks in hosts with a range in $B/T$ -- placing, e.g., M31 and M33 on more similar footing -- quenched disks exist at all $\Mstel$ which cannot be brought onto (some variant of) the \MS.  Other factors -- bars, disk dynamics, halo heating, AGN activity, environment -- must help pull these systems off the (flat) ridge-line defined by normal disks; the key point is that these processes may manifest themselves in the dispersion and not the slope of the \MS. 
 
\section{Summary}
 
Re-normalizing $\sfr$ by {\it disk} stellar mass, $\Mdisk$ can account for $\sim0.25$ dex of declining $\ssfr$ per decade $\Mstel$, essentially removing the dependence of SF efficiency on galaxy mass for star-forming systems with blue disks (if not all star-forming galaxies).  Besides suggesting a key part of ``mass-quenching" is ``bulge-building" -- distinct from processes affecting SF in disks -- our findings ease tension between the \MS\ and the evolution of the stellar mass function, and reinforce two important points: 
\begin{itemize}
	\item ``Understanding galaxy evolution demands the routine bulge--disk decomposition of the giant galaxy population at all redshifts," \citep{Allen06};
	\item Dispersion in $\sfr(\Mstel)$ likely reflects real diversity in SFHs and should not be ignored.
\end{itemize}

Upcoming IFU surveys (e.g., MaNGA; \url{www.sdss3.org/future/manga.php}) may constrain intrinsic spreads in $\sfr(M_{\ast(,\rm disk)}$) and thus mechanisms shaping SFHs. Regardless, $\sfr/\Mdisk$--$\Mstel$ and $B/T$--$\Mstel$ should serve as benchmarks for future theoretical models of galaxy evolution.

%------------------------------------------------------------------------------------------------------------------------------------------
%------------------------------------------------------------------------------------------------------------------------------------------

\section*{Acknowledgements}

L.E.A. thanks  Ryan Quadri, Daniel Masters, and Sean Johnson for many helpful discussions. He dedicates this paper to Hortense Lieberthal Zera (1916-2014), a loving grandmother and stellar personality.

Funding for the Sloan Digital Sky Survey (\url{www.sdss.org}) was provided by the Alfred P. Sloan Foundation, the Participating Institutions, the National Aeronautics and Space Administration, the National Science Foundation, the U.S.\ Department of Energy, the Japanese Monbukagakusho, and the Max Planck Society.\\

%------------------------------------------------------------------------------------------------------------------------------------------
%------------------------------------------------------------------------------------------------------------------------------------------

\bibliographystyle{apj}

\small%\bibliography{/Users/labramson/lit.bib}

%------------------------------------------------------------------------------------------------------------------------------------------
%------------------------------------------------------------------------------------------------------------------------------------------

\end{document}